# Distance to massive metal body – a paradoxical parameter that regulates the intensity of the hydration of $YBa_2Cu_3O_{6.75}$

A. V. Fetisov

*Institute of Metallurgy of the Ural Branch of the Russian Academy of Sciences, Ekaterinburg, Russia*

## 1 Introduction

A fairly limited number of chemical processes are known that can be influenced by external fields. The category of processes affected by magnetic field includes [1–3]: recombination of radicals, processes in biological systems, some adsorption processes, etc. But only some biological systems and biochemical processes going in them are affected by electric and electromagnetic fields [4]. All the systems that can be influenced by external fields are near the violation of stability conditions. At the same time, the detailed mechanism of the action of external fields on them is still not entirely clear [1–4]. In most cases, however, chemical processes do not react to either magnetic or electric field. Along with this, as far as we know, there are no any mentions in literature about the impact on chemical processes of non-magnetic and non-charged bodies which do not have direct contact with the reagents.

In our previous work [5], we studied the hydration of $YBa_2Cu_3O_{6.75}$ (YBCO) which is a superconductor with a critical temperature of about 70 K. This compound is chemically active with respect to water and as a result of prolonged contact with a moist environment can chemically degrade, losing its practical performance properties [6, 7]. In [5], however, we used a previously developed method of material preparation, such one that enables hydration process in the final product to proceed at relatively low $p_{H_2O}$ (0.9–1.0 kPa), which is unrealistic for the other methods. Hydration process carried out in such way does not lead to a sensitive chemical degradation of the material even at a high degree of saturation of it with water[1]. The most surprising thing we revealed in [5] was that the hydration kinetics of YBCO was significantly influenced by the close location of the neighboring YBCO samples to each other. Meanwhile, the samples manifested very weak magnetic properties and external fields: magnetic and electric did not affect the hydration process in them up to high values of field strength. An attempt to shield one of the interacting samples with an aluminum foil jacket did not affect the result of their in-teraction. In addition, signs of the influence on YBCO of other massive bodies, which are not superconductors, were found. Thus, upon the results of the work [5] two main unresolved questions have been remaining: what is the nature of the impact of YBCO samples on each other while they are being hydrated? And whether other bodies located next to YBCO samples can indeed influence the hydration process in them?

---
[1] It is possible that there is also the saturation of YBCO with oxygen here [5] but this point needs to be proven in additional experiments.

In the present paper, we conduct a study on how a massive steel disk located in the close vicinity of YBCO affects the hydration process proceeding in this oxide at low $p_{H_2O}$. Conducting this experiment on a large number of the YBCO samples being hydrated at different distances from the disk clearly indicates a significant inhibitory effect of this factor.

## 2 Experimental details

The YBCO material for this study was prepared by the sintering method from oxides of yttrium and copper and barium carbonate. The powders were mixed in stoichiometric proportions and heat treated first at 750°C for 5 h, then at 940°C for 150 h with intermediate grindings. After that, the product was saturated with oxygen at 585°C for 5 h in flowing air. The resulting material was examined by the iodometric titration technique for the oxygen content value ($O_{6.75}$). Besides X-ray diffraction (XRD) and X-ray photoelectron spectroscopy (XPS) was employed to characterize its structure and composition. The sintered material was ground in an agate mortar and poured into thin quartz tubes with an internal diameter of 3.0 mm (samples of type I) and quartz cups with an internal diameter of 7.3 mm (samples of type II). The cups were wrapped in two layers of aluminum foil of 11 μm thickness; several small holes were made in the foil to allow air to come inside the capsules. Finally, the prepared samples were exposed to a low humidity atmosphere with $p_{H_2O} = 110 \pm 10$ Pa for three days. The last action was an imperative for the ability of the YBCO sample to be hydrated at low $p_{H_2O}$ in further experiments [5].

Each tube and cup with YBCO sample was placed into a sealed plastic bottle with moisture holding material in it. To study the effect of massive steel disk on hydration process, the bottles were placed on it as shown in Fig. 1. For this purpose we used a non-magnetic and non-charged stainless steel disk[2] with a diameter of 70 mm and a thickness of 11 mm. The precision of measuring the distance between the flange top surface and the sample bottom (distance $l$, see Fig. 1) was ±0.5 mm except for $l = 2.0 \pm 0.1$ mm. The whole structure shown in Fig. 1 was placed in a thermostated tank in which the temperature was maintained at 24.0 ± 0.2°C. The method of monitoring the development of hydration process was weighing the tubes and con-tainers with samples on an analytical balance Shimadzu AUW-120D (Japan) with a precision of $1\cdot10^{-2}$ mg.

---

[2] Flange DN40CF, OMICRON Vacuumphysics GmbH, Germany

Since the slowing down of the hydration process observed in our work could probably be caused by any uncontrolled irradiation of the samples (rather than the effect of the steel disk), for the purity of the experiment we checked the area in which the sample was located for the presence of magnetic field (with the aid of a BST-600 gaussmeter, China) as well as β and γ radiation (with an electronic personal dosimeter). No deviations from the normal background were found during all the experimental period.

Research of the structure of YBCO was performed using a diffractometer Shimadzu XRD 7000 (Japan) with a Cu K source. Certified silicon powder was utilizing Lattice parameters were determined by the least square method, using positions of 14 diffraction lines in the range $2\Theta = 20° \div 70°$. The standard deviation between the calculated and measured positions of the lines was $\overline{\Delta 2\Theta} = 0.015°$.

The X-ray photoelectron spectroscopy (XPS) study was carried out on an electron spectrometer Multiprobe Compact equipped with an energy analyzer EA-125 (Omicron, Germany). An X-ray tube with magnesium anode of 180 W was used as the X-ray source. The spectral data acquisition was made with a step of 0.5 eV, an acquisition time of 2 s per step, and a passing energy of 100 eV. The element concentrations were calculated according to the equation:

$$X_i = \frac{I_i \sigma_i \varphi(E_b)}{\sum_i I_i \sigma_i \varphi(E_b)}, \qquad (1)$$

where $X_i$ is the $i$-th element fraction in the surface layer studied; $\sigma_i$ is the photoionization cross-section for the $i$-th element [8]; $\varphi(E_b)$ is a factor depending on the binding energy, which is determined while calibrating the spectrometer using standard samples. It is considered that if these standards are used then XPS quantitative measurements are as accurate as ±5 %. The depth of the XPS analysis is considered to be 5–10 nm.

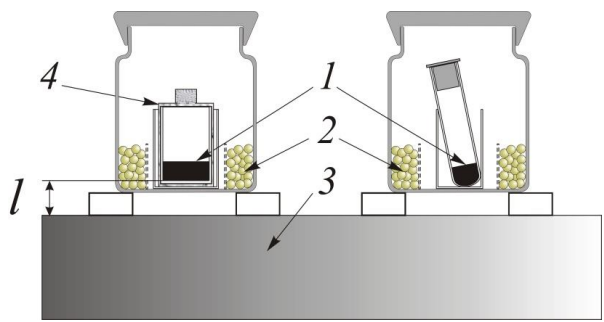

*1 – YBCO sample; 2 – moisture holding material; 3 – stainless steel disk; 4 – aluminum wrapper*

**Fig. 1.** Scheme of the experiment on the influence of massive steel disk on the development of the hydration process in YBCO.

# 3 Experimental results

## 3.1 The hydration kinetics of YBCO

Fig. 2 shows examples of the hydration kinetics of the dispersed YBCO samples placed in thin tubes (samples of type I) and cups (samples of type II). On the basis of a plenty of kinetic curves obtained by us, which are of the same type as depicted in Fig. 2, it can be stated that the hydration process under investigation proceeds equally for all samples (of type I and II) and it is characterized by reproducibility. Further, we will characterize this process with a parameter "relative hydration rate" (RHR), which will allow comparing numerically its kinetics for various YBCO samples exposed to the influence of external factors being studied in the work: RHR = ($\alpha$ / $\alpha_{max}$) · 100%, where $\alpha$ is the degree of moisture-saturation of a sample for 4 days, and $\alpha_{max}$ is the degree of moisture-saturation for 4 days of the control sample that is not subject to external factors.

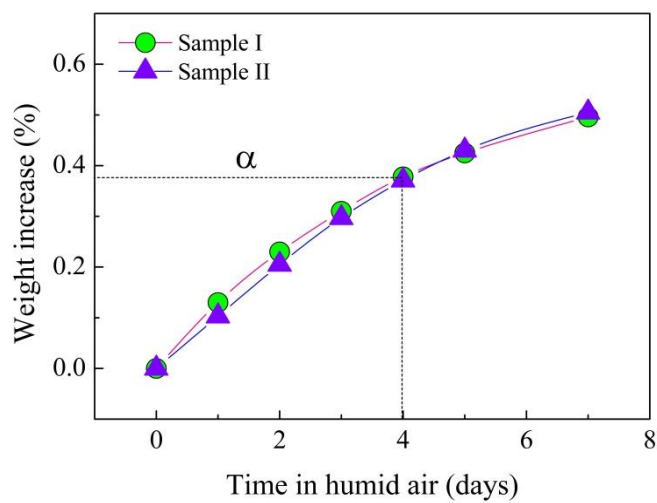

**Fig. 2.** The hydration kinetics of the YBCO samples at a humidity of 43% and a temperature of 24°C. It is shown the method of determination of the parameter $\alpha$.

## 3.2 Characterization of the initial and hydrated YBCO by XRD and XPS

The structural data and phase composition of the YBCO material at different stages of its treatment are presented in Table 1. One can state that the YBCO phase is rather contracted after exposure to air. The lattice parameters of exposed YBCO correspond to the maximally oxidized state of YBCO [9]. On the other hand, our study has not revealed radical changes in the structure that would be evidence of the transformation of YBCO under the action of absorbed water into the pseudo-$YBa_2Cu_4O_8$ phase, as had been observed in [10], or into another phase. There is ob-served only some change in the phase composition caused by the deposition of the impurity phases.

Results of the XPS measurements on the samples before and after exposure to air are depicted in Fig. 3. XPS survey spectra (Fig. 3a) show signals of all characteristic electronic levels of the elements that constitute the samples under study. A quantitative assessment of the ratio between these elements, performed according to equation (1), is presented in Table 2.

In [11], the $RBa_2Cu_3O_{6+\delta}$ (R = Y, Nd) oxides have been studied by XPS and a detailed analysis of high-resolution XPS spectra concerning these compounds has been performed. According to this work, it is convenient to evaluate the phase composition of the YBCO surface layer when considering the Ba $3d$ spectra (which are $3d_{3/2}$–$3d_{1/2}$ doublets due to spin-orbital splitting), since this electronic level is characterized by large chemical shifts. High-resolution spectra of copper and yttrium are less informative from this point of view. In [11], the Ba $3d$ binding energy in various phases containing Y, Ba, and Cu has been determined. Based on these data, to peaks at 777.6, 779.8, and 781.7 eV in the Ba $3d$ spectra obtained by us (see Fig. 3b) we put into correspondence the phases YBCO, $BaCO_3$, and $Y_2BaCuO_5$/$Ba_xCu_yO_{x+y}$ (the Ba $3d$ binding energies in the $Y_2BaCuO_5$ and $Ba_xCu_yO_{x+y}$ phases are about the same).

**Table 1.** The structural data and phase composition of the YBCO samples before and after exposure to humid air

| Structural parameter | Before exposure to air | After exposure to air |
|---|---|---|
| Major phase, space group *Pmmm* | | |
| $a$, Å | 3.8284(1) | 3.8270(1) |
| $b$, Å | 3.8884(2) | 3.8833(1) |
| $c$, Å | 11.7063(5) | 11.6931(3) |
| $V$, Å | 174.264(7) | 173.776(6) |
| Impurities | $Y_2BaCuO_5$ (~3%) | $Y_2BaCuO_5$ (~3%) |
| | | $Ba_2Cu_3O_{5.9}$ (~2%) |
| | | $BaCuO_{2.36}$ (~2%) |

Analyzing the data shown in Fig. 3 and Table 2, we can conclude that during the hydration process the material surface layer becomes depleted in yttrium atoms and enriched in copper. Shares of barium and oxygen vary slightly. Such changes can only be explained by the ap-pearance in the chemical system of a large volume of copper-containing phases, which do not contain yttrium, i.e. $Ba_xCu_yO_{x+y}$. The relative volume of these phases must be large enough to compensate in excess for the disappearance in the hydrated sample of the main phase, in which the yttrium content is small (this disappearance has arisen because of the impurity layer thickness exceeded up the thickness of the layer analyzed by XPS: 5–10 nm). This is completely consistent with the XRD data (see Table 1) according to which about 4 vol.% of $Ba_xCu_yO_{x+y}$ emerges in YBCO in the course of its hydration.

It should be noted that the formation of the $Ba_xCu_yO_{x+y}$ phases is not characteristic of the hydration of YBCO, but is characteristic of that for $NdBa_2Cu_3O_{6+\delta}$ (NBCO) [11]. As suggested in [11], this difference is associated with faster water molecule diffusion in the NBCO structure compared with their diffusion in YBCO. We believe that the appearance in the present study of the $Ba_xCu_yO_{x+y}$ phases as the hydrolysis product is precisely due to the high diffusion rate of $H_2O$ (see Fig. 2) in the YBCO material prepared according to the new method (see paragraph Experiment details).

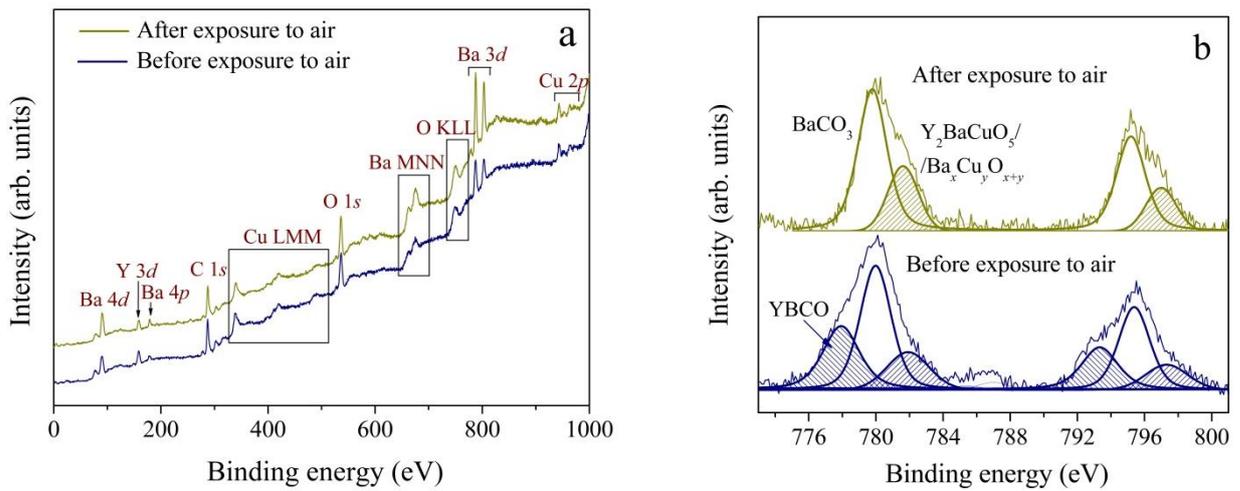

**Fig. 3. a.** The survey spectra of the YBCO samples. **b.** The Ba $3d$ spectra of the YBCO samples.

**Table 2.** The elemental composition of the YBCO sample surface layer before and after exposure to air

| Element | Before exposure to air | After exposure to air |
|---|---|---|
| Y | 0.196 | 0.042 |
| Ba | 0.115 | 0.081 |
| Cu | 0.122 | 0.188 |
| O | 0.567 | 0.689 |

## 3.3 The effect of a massive steel disk on the hydration kinetics of YBCO

Figs. 4 and 5 show the complete array of experimental data on the effect of the massive steel disk on the hydration of YBCO, accumulated by us for a long period of time and represented here in the form of RHR(W) dependences. An interesting feature of the hydration process, which has been already noted by us in the previous work [5], is striking – for a small sample weight (less than a certain value of $W_{thr}$), the process is practically not implemented. As seen from Figs. 4 and 5, the values of the parameter $W_{thr}$ for samples with diameters 3.0 and 7.3 mm are 63±4 and 305±5 mg, respectively. Taking into account the convex bottom of the samples of small diameter[3], which should lead to somewhat overestimated values of $W_{thr}$, one can consider that $W_{thr} \propto d^2$, where $d$ is the sample diameter. In other words, the critical value of W is proportional to the area of the sample base. Therefore, the value of $W_{thr}$ is achieved for all samples with a certain height $h_{thr}$. For the dispersed YBCO material used in our work, which has the bulk density ρ = 3.7 mg·mm$^{-3}$, it is easy to obtain that $h_{thr}$ = 2.1±0.2 mm. This result matches the direct measurements $h_{thr}$, which we were carrying out during the research. One, however, refutes the assumption we made earlier [5] that $h_{thr}$ depends on $d$. The critical value of W does not seem to depend on the presence of the massive steel disk near the samples. At the same time, the influence of the disk on the behavior of the RHR (W) dependences is quite noticeable at $W > W_{thr}$.

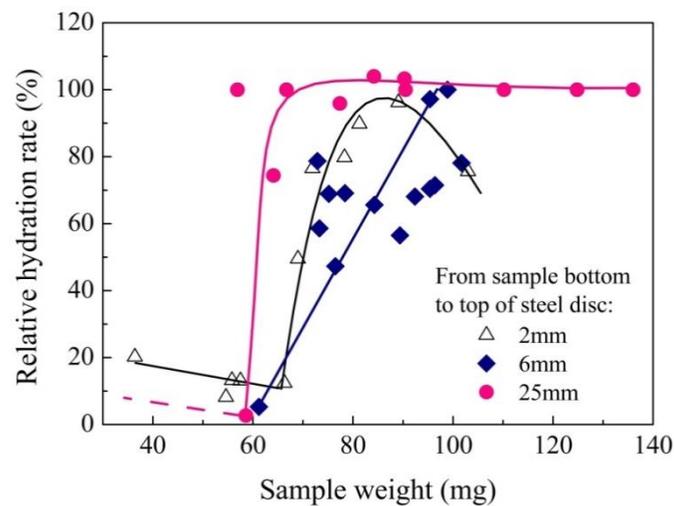

**Fig. 4.** Dependences of the relative rate of the hydration of YBCO (relative to the maximum rate for a given batch of samples) on the weight of the sample of type I.

---

[3] Due to the corresponding features of the bottom of tubes in which these samples were located.

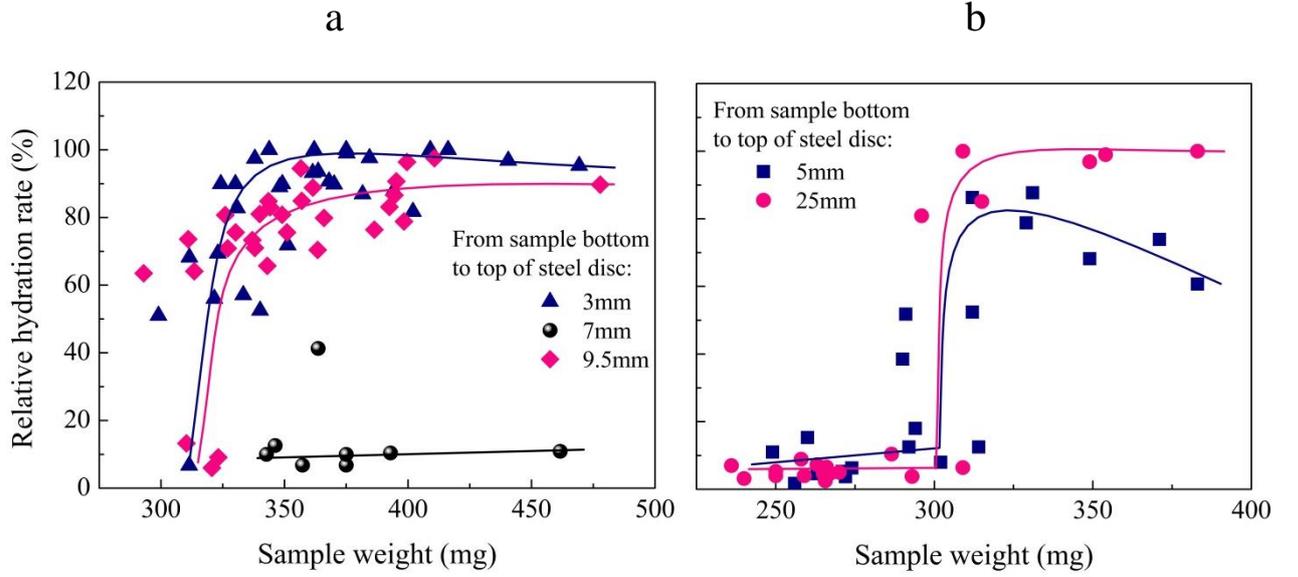

**Fig. 5.** Dependences of the relative rate of the hydration of YBCO (relative to the maximum rate for a given batch of samples) on the weight of the sample of type II.

As it follows from Figs. 4 and 5, when the steel disk is located far from a sample ($l = 25$ mm), then the increase in the hydration intensity at $W = W_{thr}$ occurs abruptly to maximum value and at $W > W_{thr}$ the hydration kinetics remains at the highest level regardless of W, which is true for any type of samples. We will consider such behavior as free from the influence of metallic body. As the steel disk approaches the sample, its inhibitory effect on the hydration kinetics of YBCO first increases, and then, after passing through a maximum at $l_{max} = 5 \div 6$ mm, it begins to decrease. In this case, the steel-disk-effect itself is manifested differently for various values of $l$. For $l \leq l_{max}$, a sharp increase in the RHR(W) dependence at $W_{thr}$ after which gradually decrease in it is characteristic. For $l \geq l_{max}$, the delayed hydration kinetics is revealed already at $W_{thr}$.

## 4  Discussion

The results obtained unambiguously indicate the existence of some influence from the side of the steel disk on the hydration of YBCO. Moreover, this impact is extreme, depending on the distance to the disk. As for the difference in the shape of the RHR(W) dependences for cases $l \geq l_{max}$ and $l \leq l_{max}$, it should be keep in mind that various parts of the sample are located at different distances from the disk, experiencing, therefore, non-equal intensity of its effect. With the sample weight increasing, this difference should manifest itself more and more significantly and differ-ently, depending on where the sample is located relative to $l_{max}$. In order to visually illustrate how the shape of the RHR(W) dependences can change with changing $l$ parameter, we will try to simulate the effect of the disk on the hydration of YBCO. To do this, we introduce the parameter $Q$ denoting intensity of the impact of the steel disk on a unit volume of YBCO. We assume in the first approximation that at a unit exposure of this kind[4], i.e. when $Q = 1$ arbitrary units, for an YBCO unit volume under consideration the relative hydration rate (RHR) decreases by $x$% and in the case of $Q < 1$, the decrease of RHR is proportional to $Q$: RHR = $100 - x \cdot Q$. Here we are free to choose the functional dependence of the intensity $Q$ on the distance $r$ between the disk and the sample unit volume under consideration. Among various extreme dependences we choose the following:

$$Q(r) = \frac{1}{ar^2 - 2ar_{max}r + c}, \qquad (2)$$

as the most suitable for describing our experimental results, where $r_{max}$ is the distance to the disk at which the function $Q(r)$ shows a maximum. The values of coefficients $a$ and $c$ are 125 and 4501 for samples of type I, and are 33 and 2129 for samples of type II. The values of the parameter $r_{max}$ are 6 and 8 for the samples of type I and II, respectively. Then the relative rate of hydration is described by the equation:

$$RHR = 100 - \frac{x}{h}\int_{l}^{l+h} \frac{dr}{ar^2 - 2ar_{max}r + c}, \qquad (3)$$

in which the height of the sample $h$ can be represented as $4W/\rho\pi d^2$. The $Q(r)$ dependence is shown in Figs. 6a and 7a, and the simulated RHR(W) curves calculated on its basis with using the value of the parameter $x = 2000$%, in Figs. 6b and 7b. It can be seen that the calculated dependences are in satisfactory agreement with the experimental ones obtained for various values of $l$, but only in the final part of them. In their initial part, the experimental dependences are not described by the given calculated, however, here we are dealing with a still unexplored effect of a sharp increase in RHR at $W_{thr}$, which cannot be considered in the framework of this simple model.

The above comparison clearly demonstrates that the effect of the massive steel disk on the hydration of YBCO found in our experiments does not have a random character and can be described as the impact on YBCO of a certain potential $Q$ having a sharp maximum on some distance from the disk. Depicted in Figs. 6a and 7a $r$-dependences of $Q$ are very similar to the result of wave interference. It is fairly obvious that steel disk itself cannot create any radiation (except for resonant acoustic). Meanwhile, it can serve as a reflector of a radiation coming from YBCO. The resulting superposition of incident and reflected waves can, in principle, generate a wave structure similar to that shown in Figs. 6a and 7a. This structure may actually be somewhat different, since we used a very simplified model to simulate it.

---

[4] Here we do not take into account the nature of this impact. We only consider its inhibitory effect on the hydration of YBCO.

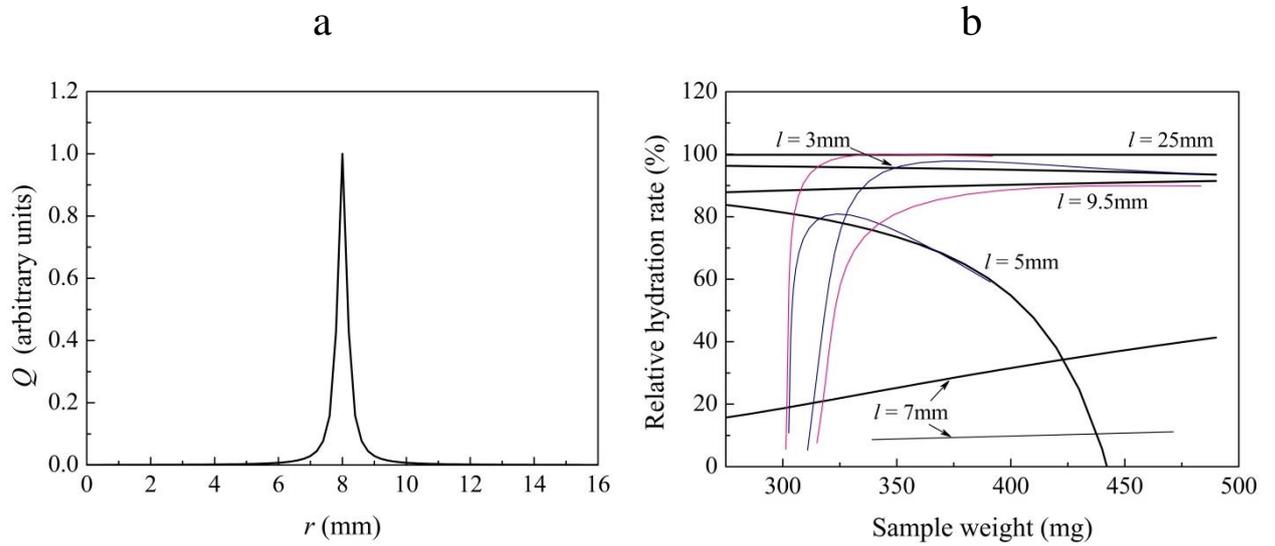

**Fig. 6. a.** The intensity of impact of the steel disk on unit volume of the YBCO sample of type II depending on the distance to the steel disk surface. **b**. Calculated dependences RHR(W), obtained with using equation (3) in comparison with experimental fitting lines in Fig. 5.

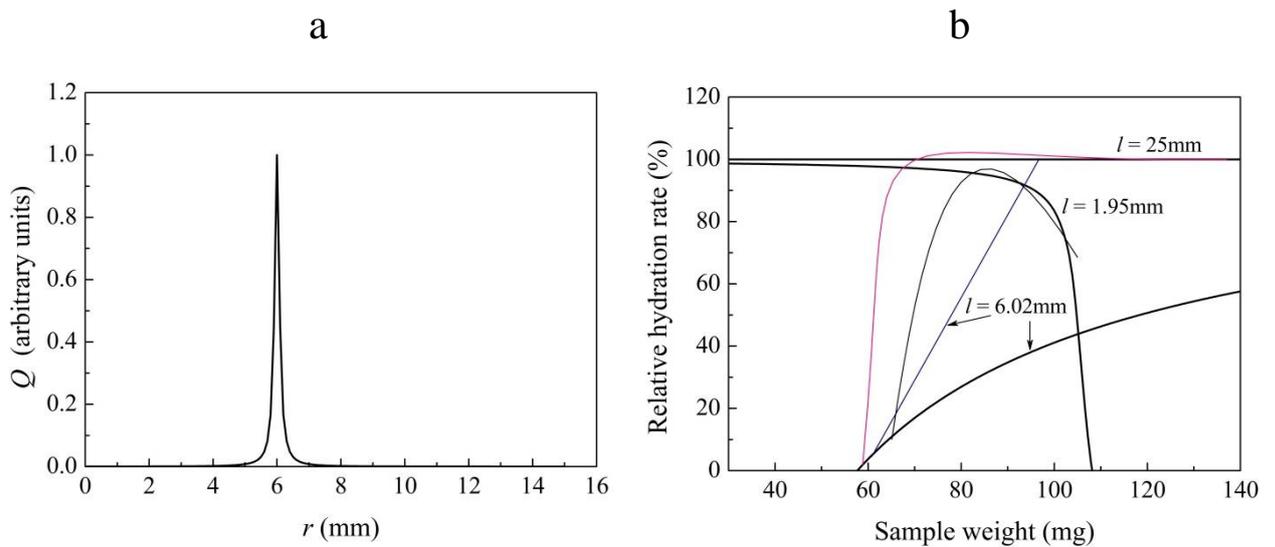

**Fig. 7. a.** The intensity of impact of the steel disk on unit volume of the YBCO sample of type I depending on the distance to the steel disk surface. **b**. Calculated dependences RHR(W), obtained with using equation (3) in comparison with experimental fitting lines in Fig. 4.

Meanwhile, the $Q(r)$ dependences obtained do not provide us with exhaustive information about the nature of the $Q$ potential. Free penetration through the aluminum screen surrounding samples of type II suggests that the radiation may be (and most likely) of electromagnetic nature[5]. Along with this, it is still difficult to explain the sustain position of the maximums in the $Q(r)$ dependences (see Figs. 6a and 7a), which are independent of the distance between disk and sample. Thus, while this question remains without a final answer.

# 5 Conclusion

Studying of the hydration of YBCO has indicated that this process is subject to the inhibitory effect of an external factor that was previously considered as not significant – the presence of a massive metal body near the sample being processed. It has been found that there is a certain distance from the metal body surface, at which the sample is exposed to maximum effect. This distance relates with the diameter of the sample. At this stage of research, it is assumed that the mediator, with the help of which the influence on the hydration process occurs, might be electromagnetic radiation emanating from a neighboring sample or reflected from the surface of metal body.

**Acknowledgments**

The author thanks collaborators of our laboratory S.Kh. Estemirova for X-ray analysis and L.A. Cherepanova for iodometric titration. The study was performed with the use of equipment of the Ural-M Collective Use Center at the Institute of Metallurgy of the Russian Academy of Sciences.

---

[5] This sort of radiation has remained after we exclude from consideration of magnetic and electric fields.